\documentclass[aps,twocolumn,nofootinbib]{revtex4}                           
\usepackage[utf8x]{inputenc}
\usepackage{epsfig}
\usepackage{amsfonts}
\usepackage{amssymb}
\usepackage{hyperref}
\usepackage{doi}
\usepackage{epsfig}
\usepackage{amsmath}
\newcommand{\beq}{\begin{equation}}
\newcommand{\eeq}{\end{equation}}
\newcommand{\kB}{k_{\mbox{\tiny B}}}
\usepackage{physics}
\usepackage{url}
\usepackage{graphicx}
\begin{document}

\title{Thermodynamics of collisional models for Brownian particles: General properties and efficiency}

\author{Angel L. L. Stable, C. E. Fern\'andez Noa, William G. C. Oropesa and C. E. Fiore}
\affiliation{Universidade de São Paulo,
Instituto de Física,
Rua do Matão, 1371, 05508-090
São Paulo, SP, Brasil}
\date{\today}
\begin{abstract}
  We introduce the idea of   {\it collisional models} for Brownian
  particles, in which
  a particle is sequentially placed in contact
  with  distinct  thermal environments and external forces.
  Thermodynamic properties
  are  exactly obtained,
  irrespective the number of reservoirs involved. In the
  presence of external forces, the
   entropy production presents a bilinear form in which Onsager coefficients are exactly calculated. Analysis of Brownian engines
  based on sequential thermal
  switchings  is proposed and  considerations about their efficiencies
  are investigated taking into account distinct external
  forces protocols. Our results shed light to a new (and alternative) route for obtaining efficient thermal engines based on finite times Brownian machines.
\end{abstract}

\maketitle

\section{Introduction}
Stochastic thermodynamics has proposed a general and unified scheme
for  addressing central issues  in thermodynamics  \cite{prigo,groot,mariobook,seifert12,broeck15}.
It includes not only an extension of  concepts from  equilibrium to nonequilibrium systems  but also it deals with the existence of new definitions and bounds  \cite{j1,j2,kau,barau},  general considerations about the efficiency of engines at finite time operations \cite{prigo,groot,mariobook} and others aspects. In all cases,
the concept of entropy production  \cite{prigo,seifert12,schn} plays a central role,
being  a quantity continuously produced
in  nonequilibrium steady states ({\text NESS}), whose main properties
and features have been extensively studied in the last years,
including {\bf its} usage for typifying phase transitions \cite{noa,esposito1,esposito2,goes}.

Basically, a {\text NESS}
can be  generated under two fundamental ways:
from fixed thermodynamic forces \cite{he,tome2015} or from time-periodic variation of external parameters \cite{seifert15,karel2016,fiorek,cleuren19}.
  In this contribution,
  we  address a
  different kind of periodic driving,  suitable for
   the  description 
 of engineered reservoirs, at which 
a system interacts  sequentially and repeatedly with
 distinct  environments  \cite{m1,m2,landio}.
 Commonly referred as {\it collisional models}, they have been inspired
 by the assumption that in many cases (e.g. the original Brownian motion)
 a particle  collides  only with
  few molecules of the environment   and then the subsequent collision will occur
    with  another fraction of uncorrelated  molecules.
  Collisional models have been viewed as   more  realistic frameworks in certain cases,
encompassing not only
particles  interacting  with a small fraction of the environment, but
also 
those presenting distinct drivings  over each member of system \cite{benn1,maru,saga,parrondo} or even  species yielding a weak coupling
 with the  reservoir. More recently, they  have  been (broadly) extended for quantum
systems  for  mimicking the environment,
represented by a  weak interaction between the system and
a sequential collection of uncorrelated particles \cite{giovanetti,esposito11,landif}.

 With the above in mind, we introduce the concept of repeated interactions
 for  Brownian particles.
 More specifically,  a particle under the influence of a given external
 force   is placed in contact with a reservoir during
 the time interval  and afterwards it is replaced by
  an entirely different (and independent) set of interactions.
 Exact expressions for thermodynamic properties
are derived and the entropy production presents a bilinear
form, in which Onsager
coefficients are obtained as function of period. Considerations
about the efficiency are undertaken and a suited regime for the system
 operating as an
efficient thermal machine is investigated.

The present study sheds light for  fresh perspectives in
nonequilibrium thermodynamics, including the possibility
of experimental buildings of  heat engines based on 
Brownian dynamics \cite{r1,r2,r3,r4,r5,r6}
with sequential reservoirs. Also, they provide us
the extension and validation  of recent bounds between currents and
entropy production,  the so called thermodynamic uncertainty relations (TURs) \cite{kau,barau,barato1,ging,r7,r81,r8},
which   has aroused a recent and great  interest.

This paper is organized as follows: Secs. II and III present
the model description and its exact thermodynamic properties. In Sec. IV
we extend analysis for external forces and considerations about efficiency
are performed in Sec. V. Conclusions and perspectives are drawn in Sec. VI.

 \section{Model and Fokker-Planck equation}
 We are dealing with a Brownian particle with  mass $m$   sequentially  placed in contact with $N$ different thermal reservoirs.
  Each contact  has a duration of $\tau/N$
 and occurs during the intervals $\tau_{i-1} \le t <\tau_{i}$, where $\tau_{i}=i\tau/N$ for $i=1,..,N$, in which
the particle evolves in time according to the following Langevin equation
\begin{equation}
    m\frac{d v_i}{dt} =  - \alpha_i v_i +F_i(t)+B_{i}(t),
  \label{two_baths_mov1}
\end{equation}
where quantities $v_i$,  $\alpha_i$  and $F_i(t)$ denote the particle velocity,
the viscous constant and external force, respectively.
From now on, we shall express them in terms of reduced quantities: $\gamma_i=\alpha_i/m$ and $f_i(t)=F_i(t)/m$. The stochastic force   $\zeta_{i}(t)=B_{i}(t)/m$ accounts for the interaction between particle  and  the $i$-th environment and satisfies the properties
\begin{equation}
 \langle \zeta_{i}(t)\rangle=0,
  \label{two_baths_ruido1}
\end{equation}
and
\begin{equation}
 \langle\zeta_{i}(t)\zeta_{i^{\prime}}(t^{\prime}) \rangle= 2 \gamma_i T_i \delta_{ii^{\prime}} \delta (t - t^{\prime}),
  \label{two_baths_ruido2}
\end{equation}
respectively,  where $T_i$ is the bath temperature.  Let $P_i(v,t)$ be the  velocity probability distribution at time $t$,  its time evolution
is described by the Fokker-Planck (FP) equation \cite{mariobook,tome2010,tome2015}
\beq
\frac{\partial P_i}{\partial t} = - \frac{\partial J_i}{\partial v} -f_i(t)\frac{\partial P_i}{\partial v}, 
\label{64}
\eeq
where $J_i$ is given by 
\begin{equation}
  \quad J_i = - \gamma_i v P_i - \frac{\gamma_i \kB T_i}{m}\frac{\partial P_i}{\partial v}.
  \label{645}
  \end{equation}
It is worth mentioning that  above equations are
formally identical to  description of the overdamped harmonic oscillator
subject to the harmonic force $f_{h} = - {\bar k}x$ just by replacing $x \rightarrow v$,
${\bar k}/\alpha \rightarrow \gamma_i$,   $1/\alpha \rightarrow \gamma_i/m$.

From  the FP equation and
by performing appropriate partial integrations together
boundary conditions in which  both $P_{i}(v,t)$ and $J_i(v,t)$
vanish at extremities, 
the time variation of the energy system $U_i=\langle E_i\rangle$
in contact with the $i$-th reservoir is given by
\begin{equation}
  \frac{dU_i}{dt}=-\frac{m}{2} \int v^2\Big[ \frac{\partial J_i}{\partial v}+f_{i}(t)\frac{\partial P_i}{\partial v}\Big]dv.
  \label{rele}
\end{equation}
The right side of Eq. (\ref{rele})
can be rewritten as  $dU_i/dt = - ( {\dot W_i}+{\dot Q_i})$,
where $\dot W_i$ and ${\dot Q_i}$ denote the work per unity of time
and heat flux from the system to the environment (thermal bath)
given by
\beq
{\dot W_i} = -m \langle v_i\rangle f_i(t)\quad {\rm and} \quad {\dot Q_i} = \gamma_i( m\langle v_i^2\rangle -  \kB T_i),
\label{112}
\eeq
respectively. In the absence of external forces $\dot W_i=0$ and all heat flux comes from/goes to
the thermal bath.

By assuming the system entropy $S$  is given by
$S_i(t) = - \kB \int P_i(v,t) \ln [P_i(v,t)]  dv$ and from the
expression for $J_i$, one finds that
its time derivative is given by
\begin{equation}
  \frac{dS_i}{dt}=-\kB \int \Big(\frac{J_i}{P_i}\Big)\Big(\frac{\partial P_i}{\partial v}\Big) dv.
\end{equation}
As for the mean energy, above expression  can be rewritten in the following form
\begin{equation}
  \frac{dS_i}{dt}=\frac{m}{\gamma_iT_i} \Big( \int \frac{J_i^2}{P_i}dv+ \gamma_i\int vJ_idv\Big).
\end{equation}
 Above expression can be interpreted according
to the following form $dS_i/dt=\Pi_i(t)-\Phi_i(t)$ \cite{tome2010,tome2015},
where  the former term
 corresponds to the entropy production rate $\Pi_i(t)$ and it is strictly positive (as expected). The second term
is the the flux of entropy and can 
also be rewritten more conveniently as
\beq
\Phi_i(t) =  \frac{\dot Q_i}{T_i}=\gamma_i\Big(  \frac{m}{T_i}\langle v_i^2\rangle -  \kB\Big).
\label{14}
\eeq
    
If external forces are null and the particle
is placed in contact to a single reservoir, the probability distribution
approaches  for large times the Gibbs (equilibrium) distribution
$P_i^{eq}(v) = e^{-E/\kB T_i}/Z$, being
$E=mv^2/2$  its kinetic  energy  and
$Z$ the partition function. 
In such case, $\langle v_i^2\rangle=\kB T_i/m$
and therefore
$\Pi_{eq}=\Phi_{eq}=0$ (as expected).
Conversely,  it will evolve to a nonequilibrium
 steady state (NESS)  when  placed in contact with sequential
 and distinct reservoirs, in which  heat is dissipated and  the
 entropy is produced and hence $\Pi_{\text NESS}=\Phi_{\text NESS}>0$.

\section{Exact solution for  arbitrary set of sequential reservoirs}
From now on,
quantities will be expressed in terms of the ``reduced temperature''
$\Gamma_i=2\gamma_i\kB T_i/m$ and $k_B=1$.
Since we are dealing with a linear force
on the velocity, the   NESS   will also be 
characterized by a Gaussian probability
distribution
$P_i(v,t)=e^{-(v-\langle v_i \rangle)^{2}/2b_i(t)}/\sqrt{2\pi b_i(t)}$ in which
both mean $\langle v_i \rangle(t)$
and the variance $b_i(t)\equiv\langle v_i^2 \rangle(t) -\langle v_i \rangle^2(t)$ will be in general  time dependent. Their expressions can be
calculated from Eqs. (\ref{64}) and (\ref{645}) and read
\begin{equation}
  \frac{d}{dt}\langle v_i\rangle=-\gamma_i\langle v_i\rangle+f_i(t),
  \label{v1}
\end{equation}
and
\begin{equation}
      \frac{d}{dt} b_i(t)  = -  2 \gamma_i   b_i(t)  + \Gamma_i,
\label{v2}
\end{equation}
respectively, where appropriate partial integrations were
performed. Their solutions  are given by the following expressions
\begin{equation}
  \langle v_i\rangle(t)=e^{-\gamma_i (t-\tau_{i-1})}[v_{i-1}'+\int_{\tau_{i-1}}^te^{\gamma_i (t'-\tau_{i-1})}f_i(t')dt'],
  \label{eqvv}
\end{equation}
  and
\begin{equation}
   b_i(t)=A_{i-1}e^{-2\gamma_i (t-\tau_{i-1})}+\frac{\Gamma_i}{2\gamma_i},
  \label{eqvc}
\end{equation}
respectively, where quantities $v_{i-1}'\equiv \langle v_i\rangle(\tau_{i-1})$ and  $A_i$'s are
evaluated by taking into account  the set of continuity relations
for the averages and variances,
$\langle v_i\rangle(\tau_i)=\langle v_{i+1}\rangle(\tau_i)$ and
$b_i(\tau_i)=b_{i+1}(\tau_i)$ (for all $i=1,...,N$),  
respectively. Since the system returns to the initial
state after a complete period, $\langle v_1\rangle(0)=\langle v_N\rangle(\tau)$ and
$b_1(0)=b_N(\tau)$, 
all coefficients  can be solely calculated
in terms of model parameters, temperature reservoirs and the period.
Also, above conditions state that
the probability  at each point
returns to the same value after every period.

 For simplicity, from now on
we shall assume
the same viscous constant $\gamma_i=\gamma$ for all $i$'s.
In the absence of external forces, all $v_{i}'$'s vanish and
the entropy production
only depends on the coefficients $A_i$'s and $\Gamma_i$'s. 
Hence, the coefficient $A_i$ becomes
\begin{equation}\label{recurrence}
A_{i+1} = xA_{i}+\dfrac{1}{2\gamma}(\Gamma_{i}-\Gamma_{i+1}),
\end{equation}
where $x=e^{-2\gamma\tau/N}$ and  all of them can be found  from
a  linear recurrence relation
\begin{equation}\label{solve}
A_{i}=x^{i-1}A_{1}+\dfrac{1}{2\gamma}\sum^{i}_{l=2}x^{i-l}(\Gamma_{l-1}-\Gamma_{l}),
\end{equation}
for  $i=2,....N$. As
the particle returns to the initial configuration
the  after a complete period, $A_N$ then reads
\begin{equation}\label{contor}
  A_{N} = x^{-1}A_{1} + \dfrac{x^{-1}}{2\gamma}(\Gamma_{1}-\Gamma_{N}). 
\end{equation}
By equaling Eqs. (\ref{solve}) and (\ref{contor}) for $i = N$, all
coefficients $A_i$'s can be finally
calculated and are given by
\begin{equation}\label{eqA1}
A_{1} = \dfrac{1}{2\gamma}\dfrac{x^{N}}{1-x^{N}}\sum^{N}_{l=1}x^{-l}(\Gamma_{l}-\Gamma_{l+1}),
\end{equation}
 and
\begin{multline}\label{eqAk}
A_{i}= \dfrac{1}{2\gamma}\dfrac{x^{i-1}}{1-x^{N}}\times\\\times\bigg[\sum^{i-1}_{l=1}x^{-l}(\Gamma_{l}-\Gamma_{l+1})+\sum^{N}_{l=i}x^{N-l}(\Gamma_{l}-\Gamma_{l+1})\bigg],
\end{multline}
for $i=1$ and $i>1$, respectively.  As we are focusing on the steady-state
  time-periodic regime, thermodynamic quantities  can
  be averaged over one period $\tau$. The mean entropy production then  $\overline{\Pi}$ reads
\begin{equation}
\begin{split}
\overline{\Pi} =\dfrac{1}{\tau}\sum^{N}_{i=1}\int^{\tau_{i}}_{\tau_{i-1}}\Phi_i(t)\,dt= \dfrac{ \left(1-e^{-2\gamma\tau/N}\right)}{2\gamma\tau}\sum^{N}_{i=1}\dfrac{A_{i}}{\Gamma_{i}}.
\end{split}
\end{equation}
From Eqs. (\ref{eqA1}) and  (\ref{eqAk}), it follows that
\begin{equation}
\sum^{N}_{i=1}\dfrac{A_{i}}{\Gamma_{i}} = \dfrac{x^{N}}{1-x^{N}}\sum^{N}_{i,l=1}x^{-l}\left(\dfrac{\Gamma_{i+l-1}-\Gamma_{i+l}}{\Gamma_{i}}\right),
\end{equation}
and we arrive at an expression for $\overline{\Pi}$  solely dependent
on the model parameters
\begin{equation}\label{Eq.prod2}
\overline{\Pi} = -\dfrac{N}{2\gamma\tau}\left(\dfrac{1-x}{x}\right)+\dfrac{1}{2\gamma\tau}\cdot\dfrac{x^{N-1}(1-x)^2}{1-x^{N}}\sum^{N}_{i,l=1}x^{-l}\dfrac{\Gamma_{i+l}}{\Gamma_{i}}.
\end{equation}
In order to show that $\overline{\Pi} \ge 0$, we resort to the
inequality
$\sum^{N}_{i=1}\Gamma_{i+l}/\Gamma_{i}\ge N\sqrt[N]{\prod^{N}_{i=1}\Gamma_{i+l}/\Gamma_{i}}$ for showing that 
  $\sum^{N}_{i=1}\Gamma_{i+l}/\Gamma_{i}\ge N$, and hence
 Eq. (\ref{Eq.prod2}) fulfills the condition
\begin{equation}
\overline{\Pi} \ge -\dfrac{N}{2\gamma\tau}\left(\dfrac{1-x}{x}\right)+\dfrac{N}{2\gamma\tau}\left(\dfrac{1-x}{x}\right)=0,
\end{equation}
in consistency
with the second law of thermodynamics.

As an concrete example, we derive explicit results for
the two sequential reservoirs case.
From Eqs. (\ref{eqvv}) and (\ref{eqvc}), coefficients $A_1$ and $A_2$
reduce to the following expressions
\begin{equation}
  A_1=\frac{\Gamma_2-\Gamma_1}{2\gamma}\Big(\frac{1-e^{-\gamma \tau}}{1-e^{-2\gamma \tau}}\Big)=\frac{\Gamma_2-\Gamma_1}{2\gamma}\Big(\frac{1}{1+e^{\gamma \tau}}\Big),
\end{equation}
where $A_2=-A_1$ and hence
\begin{equation}
  \Phi_1(t)=\gamma\Big(\frac{\Gamma_2-\Gamma_1}{\Gamma_1}\Big)\Big(\frac{1}{1+e^{2\gamma \tau}}\Big)e^{-2\gamma t},
\end{equation}
for $0\le t <\tau/2$ and
\begin{equation}
  \Phi_2(t)=\gamma\Big(\frac{\Gamma_1-\Gamma_2}{\Gamma_2}\Big)\Big(\frac{1}{1+e^{2\gamma \tau}}\Big)e^{-2\gamma (t-\frac{\tau}{2})},
\end{equation}
$\tau/2\le t <\tau$, respectively average mean entropy production
reads
\begin{equation}
  {\overline \Pi}=\Big[\frac{\Gamma_1\Gamma_2}{2\tau}\tanh{\left(\dfrac{\gamma\tau}{2}\right)}\Big]\Big(\frac{1}{\Gamma_1}-\frac{1}{\Gamma_2} \Big)^2.
  \label{epfr}
\end{equation}
Note that ${\overline \Pi}\ge 0$ and it vanishes when $\Gamma_1=\Gamma_2$.
In the limit of  slow ($\tau >>1$) and fast ($\tau <<1$) oscillations, ${\overline \Pi}$ approaches to
the following asymptotic expressions
\begin{equation}
  {\overline \Pi} \approx \frac{\Gamma_1\Gamma_2}{2\tau}\Big(\frac{1}{\Gamma_1}-\frac{1}{\Gamma_2} \Big)^2 \quad {\rm and} \quad \frac{\Gamma_1\Gamma_2\gamma}{4}\Big(\frac{1}{\Gamma_1}-\frac{1}{\Gamma_2} \Big)^2, 
\end{equation}
respectively and such a latter expression is independent on the
period.

  Eq. (\ref{epfr}) can be conveniently
written down as a flux-times-force expression, where the thermodynamic force attempts to the difference of temperatures of reservoirs. Given that
the viscous coefficient is the same for all switchings, the thermodynamic
force can be more conveniently expressed
in terms of difference of $\Gamma_i$'s.  More specifically, we have that
${\overline \Pi}={\cal J}_{T}f_T$, 
where $f_T=(1/\Gamma_2-1/\Gamma_1)$ and
${\cal J}_T$ can also be
rewritten as
${\cal J}_T=L_{TT}f_T$, where $L_{TT}$ is the Onsager coefficient given by
\begin{equation}
  L_{TT}=\dfrac{\Gamma_{1}\Gamma_{2}}{2\tau }\tanh{\left(\dfrac{\gamma\tau}{2}\right)}.
  \label{ltt}
\end{equation}
Note that $L_{TT}\ge 0$ (as expected).

\begin{figure}[!]
  \centering
\includegraphics{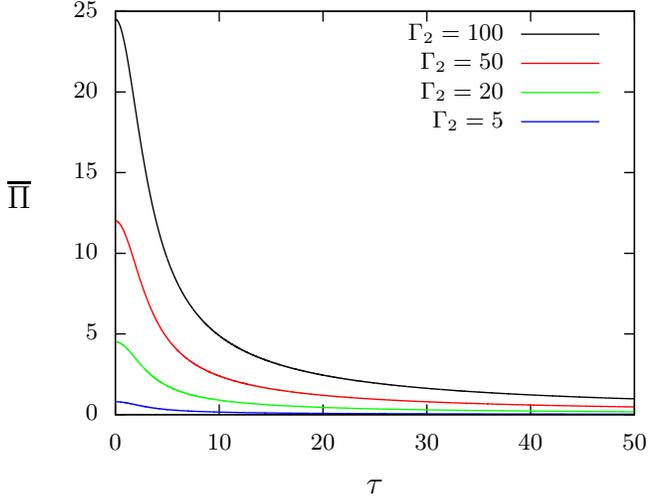}
\caption{Mean entropy production  ${\overline \Pi}$ versus  $\tau$
for distinct temperature sets $\Gamma_1=1$ and $\Gamma_2$ and $\gamma=1$.}
\label{fig1}
\end{figure}

Fig. \ref{fig1} depicts the average entropy production ${\overline \Pi}$ versus
$\tau$ for distinct values of $\Gamma_2$ and $\Gamma_1=1,\gamma=1$. Note that
it is monotonically increasing with $f_T$ and reproduces above asymptotic limits. 

\section{Forced Brownian and sequential reservoirs}
Next, we extend analysis for the case of a Brownian particle
 in contact with  sequential  reservoirs and external forces.
We shall focus on the two stage case and two
simplest  external forces protocols: constant 
and linear  drivings. More specifically,  the former
is given by
\begin{equation}
f_i(t)= \left\{\begin{matrix}
f_{1};\hspace{0.5cm}0\le t<\tau/2\\
f_{2};\hspace{0.5cm}\tau/2 \le t<\tau
\end{matrix}\right.
\end{equation}
where $f_1$ and $f_2$ denote their strengths
in the first and second half period, respectively, whereas
the latter case accounts for  forces evolving linearly over the time
according to the slopes: 
\begin{equation}
\frac{f_i(t)}{\gamma}= \left\{\begin{matrix}
 \lambda_{1}t;\hspace{0.5cm}0\le t<\tau/2\\
 \lambda_{2}(\frac{\tau}{2}-t),\hspace{0.5cm}\tau/2\le t<\tau
\end{matrix}\right.
\end{equation}
with $\lambda_1$ and $\lambda_2$ being their amplitudes. It has been considered in Ref. \cite{r8} in order to compare the performance
of distinct bounds between currents and the entropy production (TURs). 
In the presence of external forces,  FP equation
 has the same form of Eq. (\ref{eqvc}),
but now $\langle v_i\rangle(t)$'s will be different from zero.

\subsection{Constant external forces}
  From Eq. (\ref{eqvv}), the expressions for  $\langle v_i\rangle(t)$'s are given by
\begin{equation}
\langle v\rangle= \left\{\begin{matrix}
\langle v_1\rangle(t)=\dfrac{e^{\gamma\tau/2}}{\gamma}\left(\dfrac{f_{2}-f_{1}}{1+e^{\gamma\tau/2}}\right)e^{-\gamma t}+\dfrac{f_1}{\gamma},\\\\
\langle v_2\rangle(t)=\dfrac{e^{\gamma\tau/2}}{\gamma}\left(\dfrac{f_{1}-f_{2}}{1+e^{\gamma\tau/2}}\right)e^{-\gamma(t-\tau/2)}+\dfrac{f_2}{\gamma},
\end{matrix}\right.
\end{equation}
for the first or second half of each period, respectively.

The average work and heat per time  are   given by
$\overline{\dot{W}}=\overline{\dot{W}}_{1}+\overline{\dot{W}}_{2}$
and $\overline{\dot{Q}} = \overline{\dot{Q}}_{1}+\overline{\dot{Q}}_{2}$,
respectively and straightforwardly evaluated
 from Eq. (\ref{112}), whose
 $\overline{\dot{W}}_{1}$ and $\overline{\dot{Q}}_{1}$ read
\begin{equation}
\begin{split}
\overline{\dot{W}}_{1}&=-\dfrac{mf_{1}}{\tau}\int^{\tau/2}_{0}\langle v_{1}\rangle\,dt= \\
& = \dfrac{mf_{1}}{\gamma^{2}\tau}\left(f_{1}-f_{2}\right)\tanh{\left(\dfrac{\gamma\tau}{4}\right)}-\dfrac{mf^{2}_{1}}{2\gamma}, 
\end{split}
\end{equation}
 and
\begin{equation}
\begin{split}
\overline{\dot{Q}}_{1}&=\dfrac{m}{4\gamma\tau}\left(\Gamma_{2}-\Gamma_{1}\right)\tanh{\left(\dfrac{\gamma\tau}{2}\right)}+\dfrac{m}{2\gamma^{2}\tau}\left(f_1+f_2\right)^{2}\times\\&\times\tanh{\left(\dfrac{\gamma\tau}{4}\right)}+\dfrac{2mf^2_1}{\gamma^{2}\tau}\left[\dfrac{\gamma\tau}{4}-\tanh{\left(\dfrac{\gamma\tau}{4}\right)}\right],
\end{split}
\end{equation}
respectively. Analogous expressions are obtained for $\overline{\dot{W}}_{2}$ and $\overline{\dot{Q}}_{2}$ just by
exchanging $1 \leftrightarrow 2$.
Note that $\overline{\dot{Q}}_{1}+\overline{\dot{Q}}_{2}+\overline{\dot{W}}_{1}+\overline{\dot{W}}_{2}=0$, in consistency with the first law of
thermodynamics.

In the same way as before, the  steady entropy production per period $\overline{\Pi}$ can be evaluated from Eq. (\ref{14}) (by taking $\kB=1$) and reads
\begin{equation}
  \overline{\Pi}= \dfrac{2\gamma }{m}\left(\dfrac{\overline{\dot{Q}}_{1}}{\Gamma_{1}}+\dfrac{\overline{\dot{Q}}_{2}}{\Gamma_{2}}\right),
  \label{eqcc}
\end{equation}
and we arrive at the following expression
\begin{equation}
\begin{split}
  \overline{\Pi}&=\dfrac{1}{2\tau}\dfrac{\left(\Gamma_{2}-\Gamma_{1}\right)^2}{\Gamma_{1}\Gamma_{2}}\tanh{\left(\dfrac{\gamma\tau}{2}\right)}+
 \dfrac{1}{\gamma\tau}\left(\dfrac{1}{\Gamma_{1}}+\dfrac{1}{\Gamma_2}\right)\times\\  
&\times\tanh{\left(\dfrac{\gamma\tau}{4}\right)}(f_1+f_2)^2+\left(\dfrac{f^{2}_{1}}{\Gamma_{1}}+\dfrac{f^2_2}{\Gamma_{2}}\right)\left[1-\dfrac{4}{\gamma\tau}\tanh{\left(\dfrac{\gamma\tau}{4}\right)}\right].
\label{ep}
\end{split}
\end{equation}
 Since  $\gamma \tau \ge 0$ and $1-\tanh(x)/x \ge 0$, it follows   that $\overline{\Pi}\ge 0$. Note that $\overline{\Pi}$ reduces to Eq. (\ref{epfr})
as $f_1=f_2=0$.
\subsubsection{Bilinear form and Onsager coefficients}
               The shape of Eq. (\ref{ep}) is similar to the linear irreversible
thermodynamics \cite{fiorek,karel2016,proesmans2015onsager}, in which the entropy production is written
down as a sum of  flux-times-force expression. This similarity
provides to reinterpret Eq. (\ref{ep}) in the following form 
\begin{equation}
  {\overline \Pi}={\cal J}_{T}f_T+{\cal J}_1f_{1}+{\cal J}_2f_{2},
  \label{bili}
\end{equation}
 where  forces   $f_T=(1/\Gamma_1-1/\Gamma_2)$ and $f_{1(2)}$
have associated fluxes ${\cal J}_T$, ${\cal J}_{1(2)}$  given by
${\cal J}_T=L_{TT}f_T$ [identical to Eq. (\ref{ltt})],
\begin{equation}
  {\cal J}_1=L_{11}f_1+L_{12}f_2, \quad {\rm and} \quad  {\cal J}_2=L_{21}f_1+L_{22}f_2,
  \label{bili2}
\end{equation}
respectively, where $L_{11},L_{12},L_{21}$ and $L_{22}$ denote their Onsager coefficients
given by
\begin{equation}
L_{11}=\dfrac{1}{\Gamma_{1}}\left[1-\dfrac{3}{\gamma\tau}\tanh{\left(\dfrac{\gamma\tau}{4}\right)}\right]+\dfrac{1}{\gamma\tau \Gamma_2}\tanh{\left(\dfrac{\gamma\tau}{4}\right)},
\end{equation}
and
\begin{equation}
L_{12}=L_{21}=\frac{1}{\gamma\tau}\left(\dfrac{1}{\Gamma_{1}}+\frac{1}{\Gamma_{2}}\right)\tanh{\left(\dfrac{\gamma\tau}{4}\right)},
\end{equation}
respectively. Coefficients $L_{22}$ and $L_{21}$ have the same shape
of $L_{11}$ and $L_{12}$   by
replacing $1\leftrightarrow 2$, respectively. Besides, $L_{11}$ and $L_{22}\ge 0$ (as should be) and
they satisfy the inequality $4L_{11}L_{22}-(L_{12}+L_{21})^2\ge0$, in consistency
with the positivity of the entropy production.

\subsection{Time dependent external forces}
By repeating the previous calculations for  linear external forces
the mean
velocities $\langle v_i\rangle(t)$'s 
are given by
\begin{equation}
 \langle v\rangle= \left\{\begin{matrix}
 \langle v_1\rangle(t)=\frac{1}{\gamma}\bigg\{\lambda_1 \left(\gamma t-1\right)+\\\\+e^{-\gamma  t}\left[\lambda_1+   \left(\lambda_2e^{\frac{\gamma  \tau }{2}}-\lambda_1 \right)\alpha(\gamma,\tau)\right]\bigg\},\\\\ \\
 \langle v_2\rangle(t)=\frac{1}{\gamma}\bigg\{-\lambda_2\left[\gamma\left(t-\frac{\tau}{2}\right)-1\right]+\\\\+e^{-\gamma \left(t-\frac{\tau}{2}\right) }\left[   \left(\lambda_1e^{\frac{\gamma  \tau }{2}}-\lambda_2 \right)\alpha(\gamma,\tau)-\lambda_2\right]\bigg\},
 \end{matrix}\right.
 \end{equation}
 where
 $$\alpha(\gamma,\tau)=\frac{2-e^{\frac{\gamma \tau}{2}}\big(\gamma \tau-2\big)}{2(e^{\gamma  \tau }-1)},$$ respectively.
Although more complex than the previous case,
 the mean work and heat per time are evaluated analogously
 from expressions for $\langle v_i\rangle(t)$'s and $b_i(t)$'s, whose
 values averaged over a cycle read
\begin{eqnarray}
	\overline{\dot{W}}&=&-\overline{\dot{Q}}=-\mathcal{A}\bigg\{e^{\gamma  \tau }\varphi_{+}(\gamma,\tau,\xi)	\nonumber\\&+& 12e^{\frac{\gamma  \tau}{2}} \left(\gamma ^2 \tau ^2\xi  -4\right)+\varphi_{-}(\gamma,\tau,\xi)\biggr\},
\end{eqnarray}
where parameters $\mathcal{A},\xi$ and $\varphi_{\pm}(\gamma,\tau,\xi)$ read
$$
\mathcal{A}=\frac{m (\lambda_1+ \lambda_2)^{2}}{24 \gamma ^2   \tau  \left(e^{\gamma  \tau }-1\right)}
\quad\text{,}\quad
\xi=\frac{\lambda_1\lambda_2}{(\lambda_1+\lambda_2)^2},
$$
and
$$
\varphi_{\pm}(\gamma,\tau,\xi)=\gamma^2 \tau ^2 (2 \xi-1) (3\pm\gamma  \tau )+24 (1\pm\gamma \tau \xi ),
$$
respectively.
\subsubsection{Bilinear form and Onsager coefficients}
As in the previous  case, the  entropy production has also
the shape of Eqs. (\ref{bili})-(\ref{bili2}) given by
$ {\overline \Pi}={\cal J}_{T}f_T+{\cal J}_1\lambda_{1}+{\cal J}_2\lambda_{2}$,
being $L_{TT}$ the same to  Eq. (\ref{ltt}), whereas the other Onsager coefficients
read
\begin{equation}
\begin{matrix}
L_{11}=\frac{1}{\Gamma_1}\left[\frac{\gamma ^2 \tau ^2}{12}-\frac{\gamma  \tau  \left(2 e^{\gamma  \tau }+1\right)}{4 \left(e^{\gamma  \tau }-1\right)}+\frac{1}{1+e^{-\frac{\gamma  \tau }{2}}}+\frac{1}{\gamma  \tau}\tanh \left(\frac{\gamma  \tau }{4}\right)\right]+\\\\
+\frac{1}{\Gamma_2}\frac{\left[e^{\frac{\gamma  \tau }{2}} \left(\gamma  \tau -2\right)+2\right]^2}{4 \gamma \tau  \left(e^{\gamma  \tau }-1\right)},
\end{matrix}
\end{equation}
and
\begin{equation}
L_{12}=\frac{\left(2e^{\frac{\gamma  \tau }{2}}-\gamma  \tau - 2\right) \left(2e^{\frac{\gamma  \tau }{2}} -\gamma  \tau e^{\frac{\gamma  \tau }{2}}  -2\right) (\Gamma_1+\Gamma_2)}{ 4\gamma \tau  \left(e^{\gamma  \tau }-1\right) \Gamma_1 \Gamma_2 },
\end{equation}
respectively. Coefficients $L_{22}$ and $L_{21}$ are again identical to
$L_{11}$ and $L_{12}$ by exchanging $1 \leftrightarrow 2$. Also, it is straightforward
to verify that $L_{11}$ and $L_{22}$ are strictly positive and
$4L_{11}L_{22}-(L_{12}+L_{21})^2\ge0$.
\section{Efficiency}
Distinct works have tackled the conditions in  which
periodically driven systems can operate as  thermal machines \cite{seifert162,rosas16,karelc,karel2016prl,proesmans2015onsager,rosas2017stochastic}.
The conversion of a given type of energy into another one requires
 the existence of a generic force
 $X_1$  operating against its flux  $J_1X_1\le 0$ 
 counterbalancing with driving forces $X_2$ and $X_T$ in which $J_2X_2+J_TX_T\geq 0$. A measure of efficiency
$\eta$  is  given by 
\begin{eqnarray}
  \eta&=&-\frac{{\cal J}_1X_1}{{\cal J}_2X_2+{\cal J}_TX_T}\nonumber\\&=&-\frac{L_{11}X_1^2+L_{12}X_1X_2}{L_{21}X_2X_1+L_{22}X_2^2+L_{TT}X_T^2},
  \label{eta1}
\end{eqnarray}
where in such case $X_T=f_T$ and we have taken into account Eq. (\ref{bili})
for relating  fluxes and  Onsager coefficients.
 Taking into account that
 the best machine aims at maximizing the efficiency
 and minimizing the dissipation ${\overline \Pi}$ for a given
 power output 
 ${\cal P}=-\Gamma_1 {\cal J}_1X_1$,
it is important to analyze the role of three load forces, $X_{1mP}$, $X_{1mE}$
and $X_{1mS}$, in which  the power output and efficiency are maximum and
the dissipation is minimum, respectively \cite{karel2016prl}. Their values can be
obtained straightforwardly from expressions for ${\cal P}$ and Eq. (\ref{eta1}), respectively. Due to the present symmetric relation  between
Onsager coefficients $L_{12}=L_{21}$ (in both cases), they acquire simpler forms and read $2X_{1mP}=-L_{12}X_2/L_{11}$,
\begin{equation}
X_{1mE}=\frac{1}{L_{11}L_{12}X_{2}}\left[-L_{11}(L_{22}X_2^2+L_{TT}X_T^2)+A(X_2,X_T)\right],
\end{equation}
with $A(X_2,X_T)$ being given by
\begin{eqnarray}
  A(X_2,X_T)&=&\sqrt{L_{11}(L_{22}X_{2}^2+L_{TT}X_T^2)}\times\\&\times&\sqrt{[L_{11}(L_{22}X_{2}^2+L_{TT}X_T^2)-L_{12}^2X_{2}^2]}\nonumber,
    \end{eqnarray}
and $X_{1mS}=-L_{12}X_2/L_{11}=2X_{1mP}$, respectively, where
  $X_i=f_i$ and $\lambda_i$
for the constant and   linear drivings, respectively.
  The efficiencies at minimum dissipation, maximum power and its maximum value become $\eta_{mS}=0$,
\begin{equation}
\eta_{mP}=\frac{L_{12}^2X_2^2}{2(2L_{22}L_{11}-L_{12}^2)X_{2}^2+4L_{TT}L_{11}X_{TT}^2},
\end{equation}
and 
\begin{multline}  \label{eqme}
\eta_{mE}=\frac{1}{L_{12}^2X_{2}^2}[2L_{11}(L_{22}X_{2}^2+L_{TT}X_{TT}^2)-\\
-L_{12}^2X_2^2-2A(X_2,X_T)],
\end{multline}
respectively, and finally their associated  power outputs read ${\cal P}_{mS}=0$,
${\cal P}_{mP}=\Gamma_1 L_{12}^2X_2^2/4L_{11}$ and
\begin{multline}
{\cal P}_{mE}=\frac{\Gamma_1}{L_{11}L_{12}^2X_2^2}\times\\ \times \left[L_{11}(L_{22}X_2^2+L_{TT}X_T^2)-A(X_2,X_T)-L_{12}^2X_{2}^2\right]\times\\\times \left[L_{11}(L_{22}X_2^2+L_{TT}X_T^2)-A(X_2,X_T) \right],
\label{eqpme}
\end{multline}
respectively.
We pause to make a few comments:
First, above expressions extend
the findings from Ref. \cite{karel2016prl}
for  a couple of driving forces.  Second,
both efficiency and power vanish when  $X_1=X_{1mS}$ and $X_1=0$ 
and are strictly positive between those limits. Hence 
the physical regime in which the system can
operate as an engine is bounded by the lowest entropy production
 ${\overline \Pi}_{mS}=L_{TT}X_T^2+(L_{22}-L_{12}^2/L_{11})X_2^2$
and the value ${\overline \Pi}^*=L_{TT}X_T^2+L_{22}X_2^2$.
Third, despite the long expressions for Eqs. (\ref{eqme}) and (\ref{eqpme}),
powers ${\cal P}_{mP},{\cal P}_{mE}$ and efficiencies $\eta_{mP},\eta_{mE}$ are
linked through a couple of simple expressions (in similarity 
with Refs. \cite{karel2016prl,karelc}):
\begin{equation}
  \eta_{mP}=\frac{\eta_{mE}}{1+\eta^2_{mE}} \qquad {\rm and} \qquad \frac{{\cal P}_{mE}}{{\cal P}_{mP}}=1-\eta^2_{mE},
\end{equation}
and  they imply that
$0\le \eta_{mP}< \eta_{mE}$ (with $0\le \eta_{mE}\le 1$ and $0\le \eta_{mP}\le 1/2$) and $0\le {\cal P}_{mE}\le {\cal P}_{mP}$.
Fourth and last, the achievement of most efficient machine $\eta_{mE}=1$
implies that the system has to be operated at null power ${\cal P}_{mE}=0$
and hence the projection of a machine  operating for  finite ${\cal P}_{mP}/{\cal P}_{mE}$  will imply at a loss of its efficiency.

\begin{figure*}[t!]
  \centering
  \includegraphics{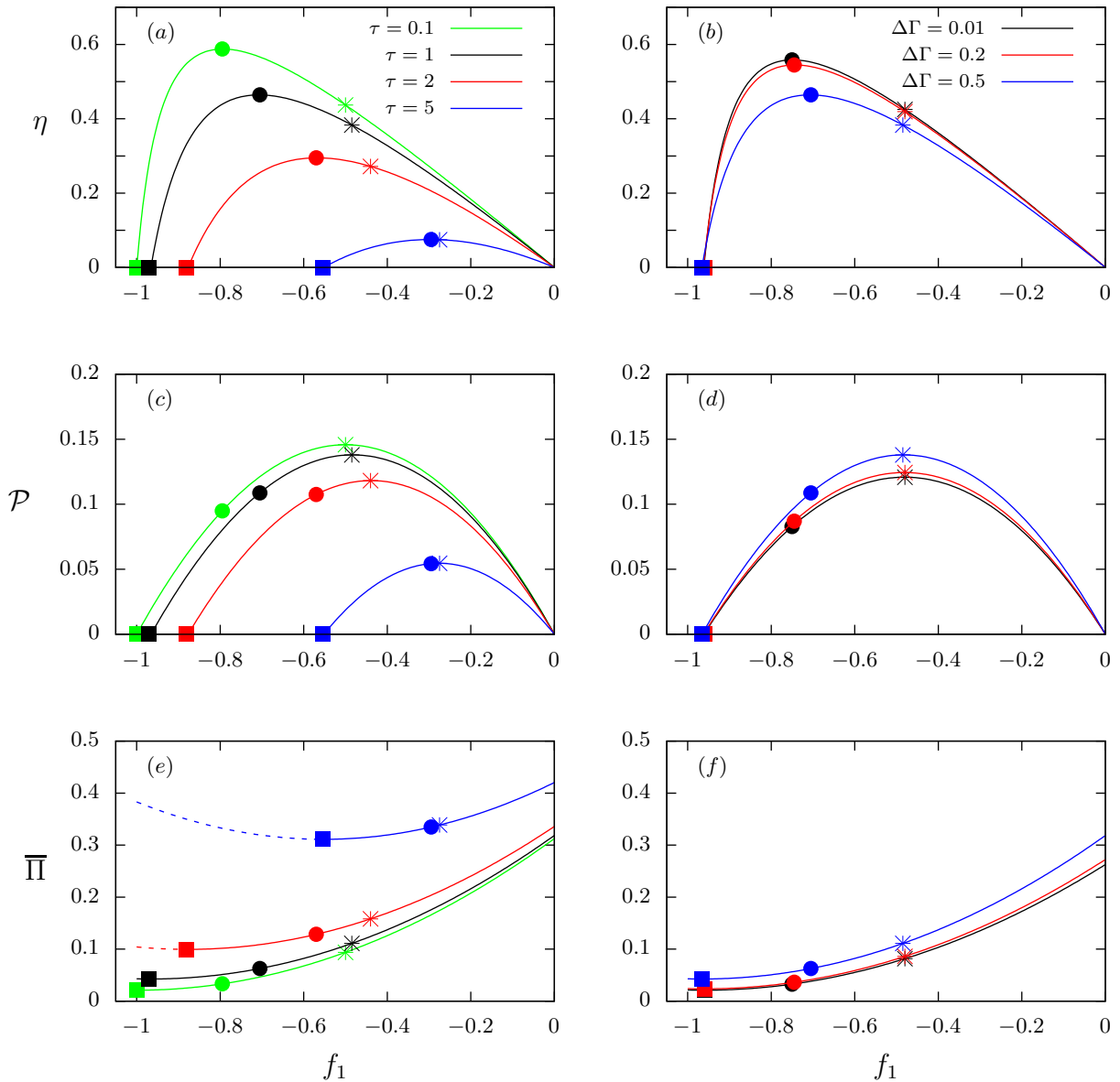}
\caption{Panels $(a)$ and $(b)$  depict the efficiency $\eta$ versus  $f_1$
  for distinct 
  periods $\tau$ (for $\Delta \Gamma=0.5$) and  $\Delta \Gamma$'s (for $\tau=1$),  respectively. In both cases, $\Gamma_1=2$ and $f_2=1$. Symbols
  $\bullet$, ``stars'' and ``squares'' denote the $f_{1mE}$, $f_{1mP}$
  and $f_{1mS}$ respectively.
  Panels $(c)$ and $(d)$ show  the
  corresponding  power ${\cal P}$, whereas $(e)$ and $(f)$ the average entropy production rate  ${\overline \Pi}$. Dashed
  lines show the values of $f_1$ the system can not be operated as a
thermal machine. }
\label{fig2}
\end{figure*}

\begin{figure*}
  \centering
  \includegraphics[scale=1]{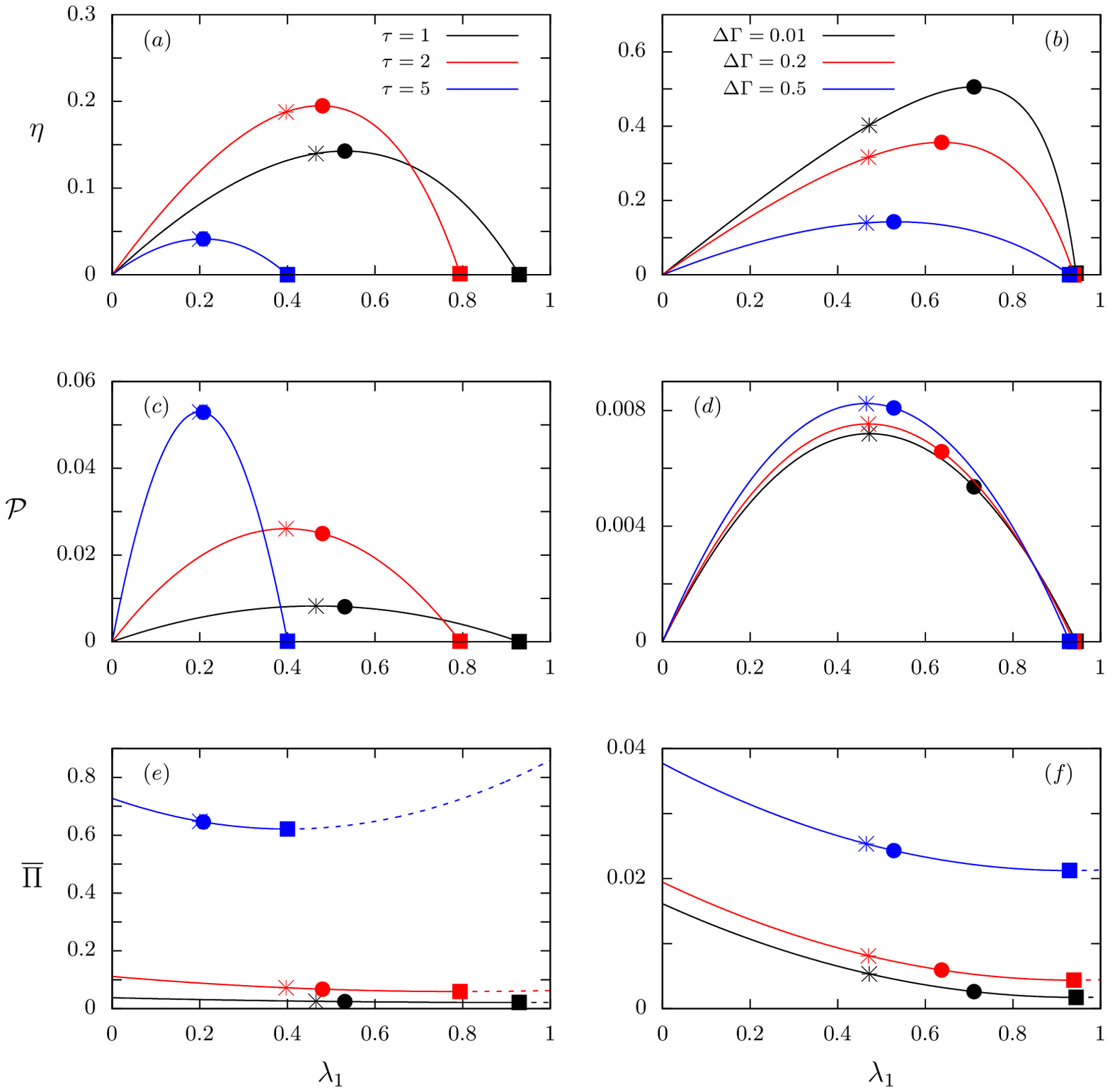}
\caption{Panels $(a)$ and $(b)$  depict the efficiency $\eta$ versus  $\lambda_1$
  for distinct 
  periods $\tau$ (for $\Delta \Gamma=0.5$) and  $\Delta \Gamma$'s (for $\tau=1$),  respectively. In both cases, $\Gamma_1=2$ and $\lambda_2=1$. Symbols
  $\bullet$, ``stars'' and ``squares'' denote the $\lambda_{1mE}$, $\lambda_{1mP}$
  and $\lambda_{1mS}$ respectively.
  Panels $(c)$ and $(d)$ show  the
  corresponding  power ${\cal P}$, whereas $(e)$ and $(f)$ the average entropy production rate  ${\overline \Pi}$. Dashed
  lines show the values of $\lambda_1$ the system can not be operated as a
thermal machine. }
\label{fig3}
\end{figure*}

Our purpose here  aims at not only extending
  relevant concepts about efficiency for Brownian particles
  in contact with sequential reservoirs, but also to show that
  a desired compromise between maximum power
  and maximum efficiency can be achieved by adjusting conveniently the model
  parameters (such as the period and the driving). From expressions
  for Onsager coefficients,
aforementioned quantities are evaluated, as depicted 
  in Figs. \ref{fig2}  and  \ref{fig3}  for distinct periods $\tau$ and temperature differences $\Delta \Gamma$'s for constant and linear drivings, respectively.
In both cases,  quantities follow  theoretical
predictions and  exhibit similar portraits, in which
efficiencies and power outputs present maximum values at $f_{1mE}(\lambda_{1mE})$ and $f_{1mP}(\lambda_{1mP})$,
respectively. The  loss of efficiency  from the maximum  $\eta_{mE}$ as $f_1$($\lambda_1$)
goes up (down)
is signed by the increase of dissipation (as expected) until
vanishing when ${\overline \Pi}={\overline \Pi}^*$. 
For the constant driving,  absolute values of forces and efficiencies
increase as the   period $\tau$ (see e.g. panels $(a)$) and/or
temperature differences
(see e.g. panels $(b)$) are lowered.
In such a case, $\Gamma_1 \approx \Gamma_2=\Gamma$,
 $\Delta \Gamma=\Gamma_1-\Gamma_2<<1$ and 
 the thermodynamic force $f_T$ approaches to $f_T \approx \Delta \Gamma/\Gamma^2$.
Onsager coefficients become simpler in
the limit of fast switchings, $\tau \rightarrow 0$ and
$L_{11},L_{22},L_{12}$  approach  to $(\Gamma_1+\Gamma_2)/(4\Gamma_1\Gamma_2)$. Some remarkable quantities  then approach to the 
asymptotic values
$f_{1mS} \rightarrow -f_2=2f_{1mP}$ and
\begin{equation}
  \eta_{mP} \rightarrow \frac{f_2^2(\Gamma_1+\Gamma_2)}{2[f_2^2(\Gamma_1+\Gamma_2)+2\Delta \Gamma^2]},
\end{equation}
respectively. For $\Gamma_1\approx \Gamma_2$,
$\eta_{mP} \rightarrow 1/2$, $\eta_{mE} \rightarrow 1$ and
${\cal P}_{mP}$  reads  ${\cal P}_{mP}\rightarrow f_2^2/8$
and thereby the limit of an ideal machine
is achieved for low periods and equal temperatures. Similar
features are verified
for the linear driving, including
 increasing efficiencies  as both $\tau$ and $\Delta \Gamma$
decreases. However, they are marked
by a reentrant behavior for $\tau<<1$ and $\Delta \Gamma \neq 0$  (see e.g. Figs. \ref{fig3}(a) and \ref{fig5}). It moves for lower $\tau$'s as
$\Delta \Gamma$ goes down and the limit of
ideal machine,
$\eta_{mP} \rightarrow 1/2$ and $\eta_{mE} \rightarrow 1$, is also recovered
when both $\tau \rightarrow 0$ for $\Delta \Gamma \rightarrow 0$.

\begin{center}
\begin{figure}
  \centering
  \includegraphics[scale=1]{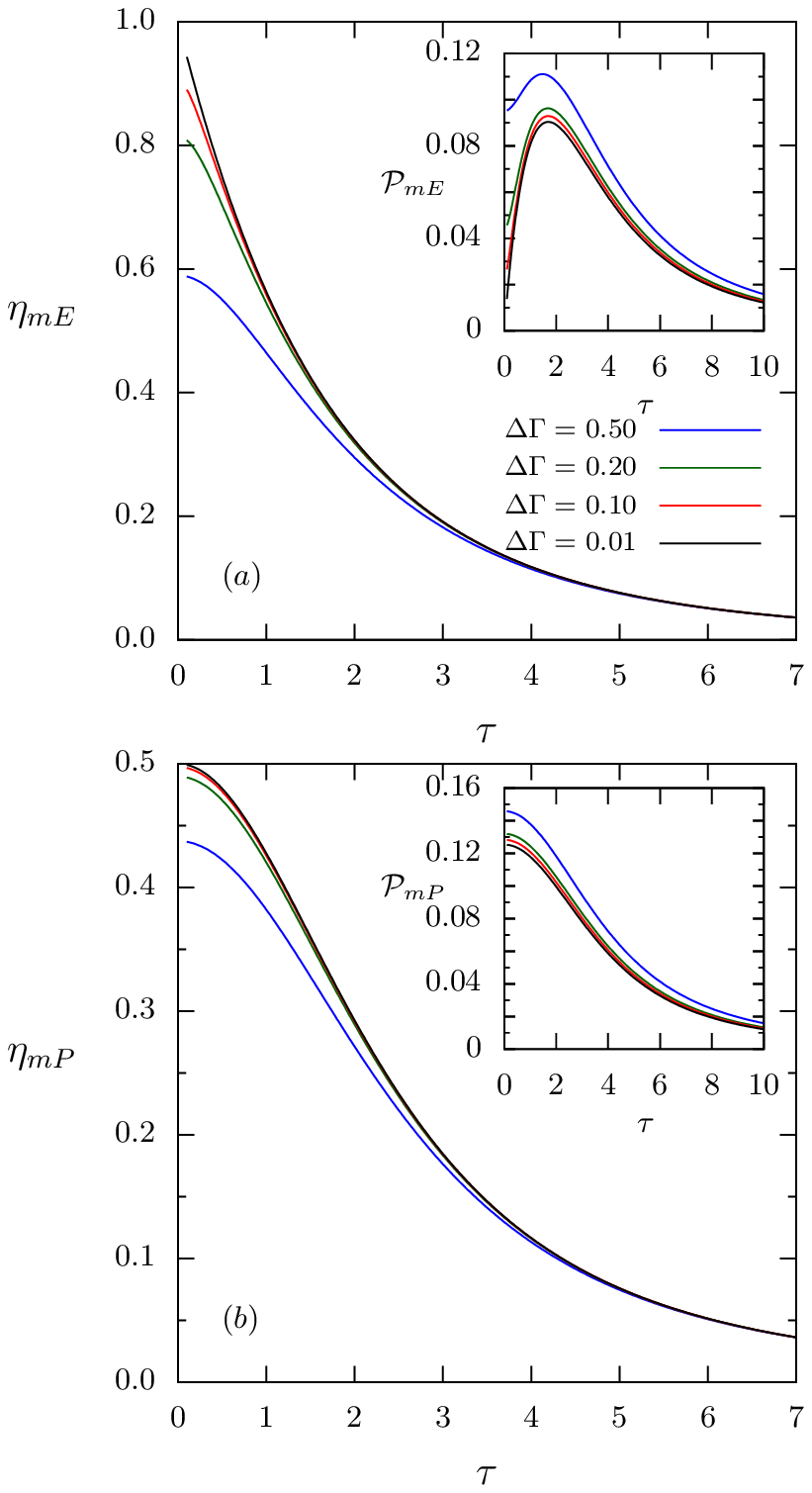}
  \caption{For $\Gamma_1=2$, $f_2=1$ and distinct $\Delta \Gamma$'s, the comparison between maximum efficiency (panel $(a)$)
    and efficiency at maximum power (panel $(b)$) for constant drivings. Insets: The corresponding
  power outputs  ${\cal P}$'s versus $\tau$.}
\label{fig4}
\end{figure}
\end{center}

\begin{center}
\begin{figure}
  \centering
  \includegraphics[scale=1]{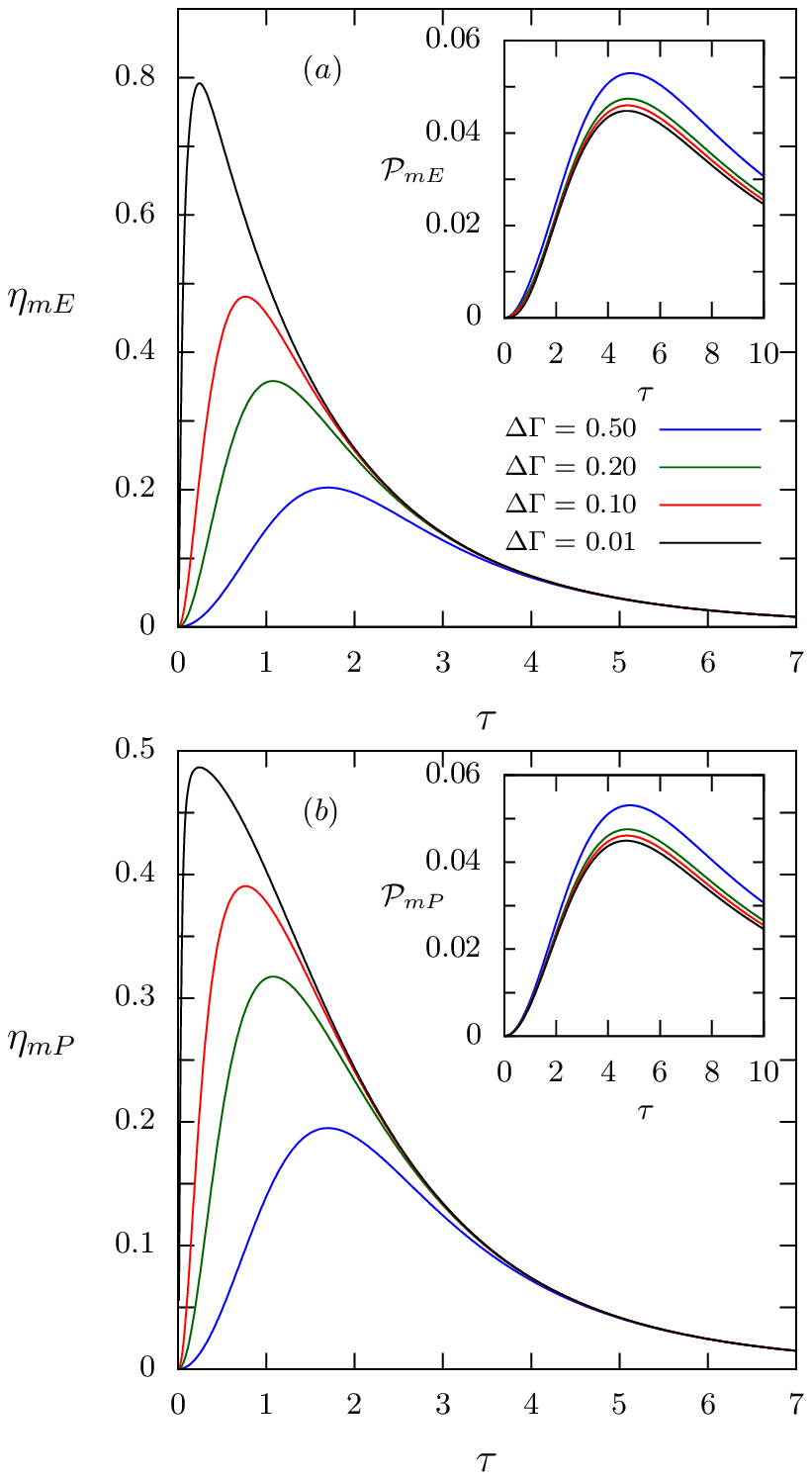}
  \caption{For $\Gamma_1=2$, $\lambda_2=1$ and distinct $\Delta \Gamma$'s, the comparison between maximum efficiency  (panel $(a)$)
    and efficiency at maximum power (panel $(b)$) for linear drivings. Insets: The corresponding
  power outputs  ${\cal P}$'s versus $\tau$.}
\label{fig5}
\end{figure}
\end{center}

Other differences between  protocols are appraised in Figs. \ref{fig4}
and \ref{fig5}.
 For finite difference of temperatures, the constant driving
  is always more  efficient than the linear one and their  power outputs
  are also superior. The maximum efficiency curves (linear
  drivings) are also reentrant, whose   maxima values increase and deviate
  for lower $\tau$'s as $\Delta \Gamma$ decreases.

We close this section by remarking that
although  short periods   indicates  a general route for optimizing
the efficiency  of thermal machines in contact to sequential reservoirs,
the present description provides to properly tune  the period and forces
in order to obtain the desirable compromise between maximum efficiency and power.


  \section{Conclusions}
  The thermodynamics of a  Brownian particle  periodically placed in contact with sequential
  thermal reservoirs is introduced.
  We have obtained explicit (exact)
  expressions for relevant quantities, such as
  heat, work and entropy production. Generalization
  for an arbitrary number of sequential reservoirs and
  the influence of external forces
  were considered. Considerations about the efficiency were undertaken,
  in which Brownian  machines can be properly operated
  ensuring the reliable
  compromise between efficiency and power  for small switching periods.

  As a final comment, we mention the several
  new perspectives to be addressed. First, it  might be very interesting
  to extend such study for other external forces protocols (e.g. sinusoidal time dependent ones)
  as well as for
  time asymmetric switchings, in order
  to compare their efficiencies, mainly with the linear driving case. Finally, it   would
 be  very remarkable to verify the validity of
 recent proposed uncertainties relations (TURs) for Fokker-Planck
  equations \cite{r7,r8}, in such class of systems.
\section{Acknowledgment}
We acknowledge Karel Proesmans and M\'ario J. de Oliveira for  careful readings of the manuscript
and useful suggestions.
C. E. F acknowledges the financial support from FAPESP under grant
  2018/02405-1.
\bibliographystyle{apsrev4-1}
\bibliography{references}
 
\end{document}